# Magnetic field dependence of the temperature derivative of resistivity: a probe for distinguishing the effects of pseudogap and superconducting fluctuations in cuprates

S. H. Naqib[a,b*], J. R. Cooper[b], R. S. Islam[a,b]

[a]*Department of Physics, Rajshahi University, Raj -6205, Bangladesh*

[b]*IRC in Superconductivity and Department of Physics, University of Cambridge, Madingley Road, Cambridge CB3 0HE, UK*

**Abstract**

We have studied the magnetic field dependence of the temperature derivative of the resistivity, $d\rho(H,T)/dT$, of a number of $Y_{1-x}Ca_xBa_2(Cu_{1-y}Zn_y)_3O_{7-\delta}$ crystalline thin films over a wide range of sample compositions. From the analysis of the temperature derivative data we have been able to distinguish quite clearly between two characteristic temperature scales, (a) the onset of strong superconducting fluctuation temperature, $T_{scf}$ and (b) the pseudogap temperature, $T_{PG}$. Significantly different characteristic features of $d\rho(H,T)/dT$ at $T_{scf}$ and at $T_{PG}$ imply that $T_{scf}$ and $T_{PG}$ have different physical origins.

## 1. Introduction

The temperature dependence of the resistivity, $\rho(T)$, provides a way to obtain the characteristic pseudogap (PG) temperature, $T_{PG}$, from the widely used gradual characteristic downturn in $\rho(T)$ at a particular temperature, much above $T_c$ in the underdoped (UD) samples [1]. As hole content, p, increases $T_{PG}$ falls sharply. When $p \geq p_{opt}$, (at $p_{opt}$ superconducting (SC) transition temperature, $T_c$, is maximum) it becomes hard to distinguish between an accelerating downturn in $\rho(T)$ near $T_c$ due to strong SC fluctuations and the one due to PG. In this study magnetic field (H) dependent temperature derivative of resistivity, $d\rho(H,T)/dT$, has been used to distinguish between the effects of PG and strong SC fluctuations near $T_c$. We have been able to locate a characteristic temperature, $T_{scf}$, near $T_c$ where $\rho(T)$ starts a rapid and accelerating downturn leading to SC transition. $T_{scf}$ is the onset temperature for strong SC fluctuations. Significantly different Zn- and H-dependences of $d\rho(H,T)/dT$ at $T_{scf}$ and at $T_{PG}$ imply that

*Corresponding author. Tel.: +88-(0)721-750288; Fax: +88-(0)721-750064; e-mail:salehnaqib@yahoo.com

$T_{scf}$ and $T_{PG}$ have different physical origins.

## 2. Experimental samples results

High-quality c-axis oriented thin films of $Y_{1-x}Ca_xBa_2(Cu_{1-y}Zn_y)_3O_{7-\delta}$ were grown on (*100*) SrTiO$_3$ substrates using the pulsed laser deposition (PLD). Details of the PLD parameters and characterization of the films can be found in ref.[2]. The p-values were determined from the room-T thermopower [2,3] and were changed by annealing the films at different T and O$_2$ partial pressures [2]. Details of $\rho(H,T)$ measurements can be found in ref.[4]. We have measured the *ab*-plane resistivity for a total of twelve films. Representative $\rho(H,T)$ data are shown in Fig. 1 and $d\rho(H,T)/dT$ for the same samples in Fig. 2. Note that $T_{scf}$ has been defined as the temperature where $d\rho(H,T)/dT$ becomes H-dependent, and that $T_{PG}$ is completely H-independent, at least up to 12 tesla. The effect of Zn is to suppress both $T_c$ and $T_{scf}$, but it does not change $T_{PG}$, while Ca only changes $T_{PG}$ via p [2,4]. Fig. 2a and 2c show that $T_c$ and $T_{scf}$ of the slightly UD (p = 0.15) $Y_{0.95}Ca_{0.05}Ba_2Cu_3O_{7-\delta}$ are 85K and 102K respectively, while they are 64.5K and 81K for the 2%Zn substituted, optimally doped (p = 0.164) sample. 2%Zn suppresses $T_c$ and $T_{scf}$ almost by the same amount from those for Zn-free compounds at the same p. $T_{PG}$ for this 2%Zn film is 111K (same as that for pure Y123 [4] with identical p). Fig. 3 shows $d\rho(H,T)/dT$ for an overdoped (OD) $Y_{0.95}Ca_{0.05}Ba_2(Cu_{0.98}Zn_{0.02})_3O_{7-\delta}$ with p = 0.174 and $T_c$ = 63.5K. Here, $T_{PG}$ and $T_{scf}$ are quite close together, 90K and 79K, respectively. It is worth mentioning that $T_c$ for pure Y123 is 90K at this hole content [2,4].

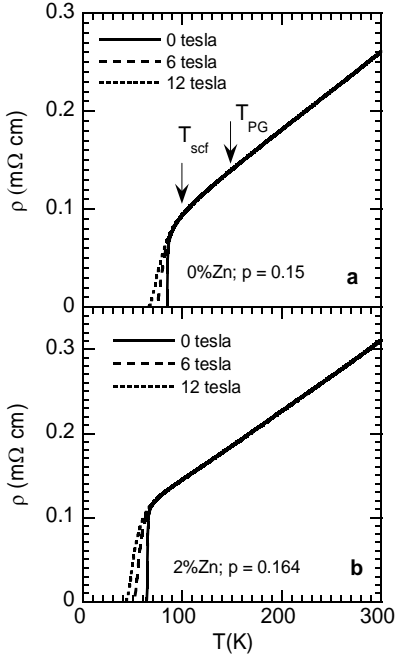

Fig. 1. $\rho(H,T)$ for (a) $Y_{0.95}Ca_{0.05}Ba_2Cu_3O_{7-\delta}$, $T_{PG}$ and $T_{scf}$ are shown (for details see text and Fig. 2) and (b) $Y_{0.95}Ca_{0.05}Ba_2(Cu_{0.98}Zn_{0.02})_3O_{7-\delta}$.

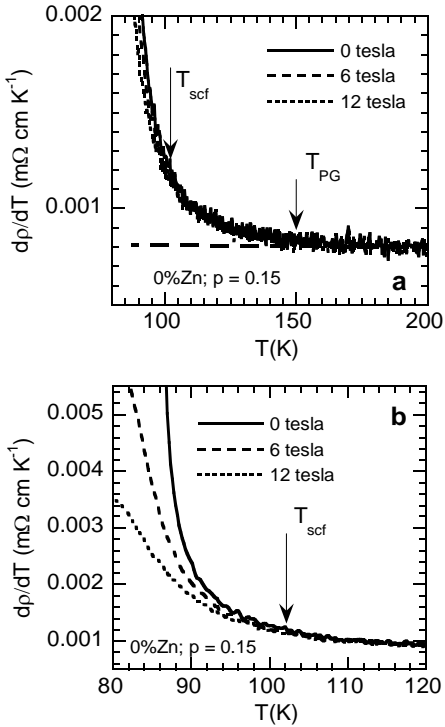

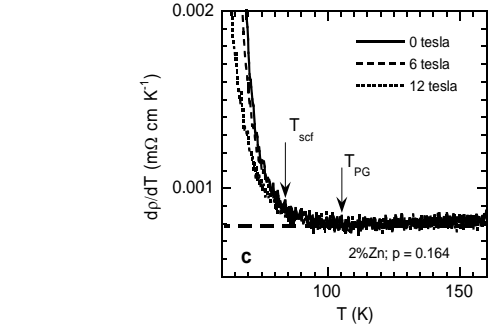

Fig. 2. Identification of $T_{PG}$ and $T_{scf}$ for (a) $Y_{0.95}Ca_{0.05}Ba_2Cu_3O_{7-\delta}$ (p = 0.15), (b) $T_{scf}$ is shown more clearly here for $Y_{0.95}Ca_{0.05}Ba_2Cu_3O_{7-\delta}$ (p = 0.15) and (c) $Y_{0.95}Ca_{0.05}Ba_2(Cu_{0.98}Zn_{0.02})_3O_{7-\delta}$ (p = 0.164).

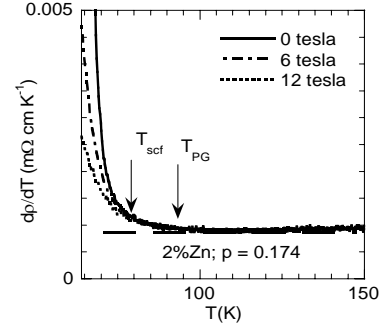

Fig. 3. $T_{PG}$ and $T_{scf}$ of an OD (p = 0.174) $Y_{0.95}Ca_{0.05}Ba_2(Cu_{0.98}Zn_{0.02})_3O_{7-\delta}$.

## 3. Discussions and conclusion

In this study $d\rho(H,T)/dT$ was used to distinguish between the two characteristic temperatures, $T_{PG}$ and $T_{scf}$. $T_{PG}$ was identified by the usual method (at $T_{PG}$, $\rho(T)$ starts falling below its high-T linear behaviour) [1,4]. $T_{scf}$ was taken at the temperature where $d\rho(H,T)/dT$ becomes field-dependent. Previously, based on SC fluctuation analysis, we identified $T_{scf}$ from the $\rho(T)$ data in zero magnetic field [4]. The present $T_{scf}(p)$-values are in excellent agreement with those found earlier[4], and again $T_{scf}(p)$ and $T_c(p)$ show identical p-dependences [4]. Strikingly different p-, Zn-, and H-dependences of $T_{PG}$ and $T_{scf}$ point towards a non-SC origin of the PG.

### Acknowledgements

Authors thank Dr. J. W. Loram and Prof. J. L. Tallon and for helpful suggestions. SHN thanks members of the Quantum Matter group, Department of Physics, University of Cambridge, UK, for their hospitality.